\documentclass[aps,prl,twocolumn,superscriptaddress]{revtex4-1}

\usepackage{graphics}
\usepackage{graphicx}
\usepackage{color}
\usepackage{epstopdf}
\usepackage{amsmath}
\usepackage{amssymb}
\usepackage{ wasysym }
\usepackage{xcolor}
\usepackage{fdsymbol}

\usepackage{cancel} 

\begin{document}

\title{Droplets capped with an elastic film can be round, elliptical, or nearly square} 

\author{Rafael D. Schulman}
\affiliation{Department of Physics and Astronomy, McMaster University, 1280 Main St. W, Hamilton, ON, L8S 4M1, Canada.}
\author{Kari Dalnoki-Veress}
\email{dalnoki@mcmaster.ca}
\affiliation{Department of Physics and Astronomy, McMaster University, 1280 Main St. W, Hamilton, ON, L8S 4M1, Canada.}
\affiliation{Laboratoire de Physico-Chimie Th\'eorique, UMR CNRS Gulliver 7083, ESPCI Paris, PSL Research University, 75005 Paris, France.}

\date{\today}

\begin{abstract}

We present experiments which show that the partial wetting of droplets capped by taut elastic films is highly tunable. Adjusting the tension allows the contact angle and droplet morphology to be controlled. By exploiting these elastic boundaries, droplets can be made elliptical, with an adjustable aspect ratio, and can even be transformed into a nearly square shape. This system can be used to create tunable liquid lenses, and moreover, presents a unique approach to liquid patterning.

\end{abstract}

\pacs{}

\maketitle

Wetting has been the subject of intense research for well over a century, motivated largely by industrial applications, ranging from designing tires to treating textiles and choosing coatings for surfaces, but also by the academic interest in a field which is undeniably rich~\cite{Gennes2008}. A common theme in wetting phenomena is that boundary conditions and substrate play a pivotal role. For instance, when the substrate is replaced by a soft solid, elastocapillary interactions lead to deviations from the classical Young-Dupr\'{e}'s law due to the formation of a wetting ridge at the contact line~\cite{Best2008,Jerison2011,Style2012,Style2013a,Marchand2012a,Park2014,Hui2014,Bostwick2014}. The substrate can also be replaced by a thin free-standing elastic film, serving as a compliant boundary for the droplet~\cite{Shanahan85,Nadermann2013,Hui2015,Hui2015b,Schulman2016,Liu2016,Fortais2017,Schulman2017b}. In this geometry, the contact angles are set by a Neumann construction with mechanical and interfacial tensions balanced at the contact line~\cite{Nadermann2013,Hui2015,Schulman2016,Fortais2017,Schulman2017b}.

Compliant elastic surfaces also show novel wetting behaviours and morphologies. For instance, droplets have been observed to migrate towards regions of increased compliance~\cite{Style2013,Alvarez2018,Liu2017}, and interact with other droplets due to deformations induced in the elastic films~\cite{Karpitschka2016,Liu2016}. Dewetting liquid films capped by a thin elastic layer can have their dynamics and dewetting morphologies controlled by adjusting the tension~\cite{Schulman2018b}. Anisotropic tension in a supporting free-standing film causes sessile droplets to elongate along the high tension direction, and thus, droplets map out the stress field in the elastomer~\cite{Schulman2017b}.  Furthermore, droplets pressed between a rigid surface and a soft solid acquire the shape of a flattened ellipsoid~\cite{Martin1997}. 

Although partial wetting on soft or compliant solids has received significant attention, here, we examine partial wetting in a novel geometry wherein droplets are capped by a thin elastic film under tension. This system could serve as a model for blisters or droplets trapped beneath drying paint~\cite{Farinha2000}. We show that the contact angle of these droplets decreases with increased tension which  is well described through an analogy made with Young-Dupr\'{e}'s law which incorporates mechanical tension. The model contains a free parameter from which the elastomer-liquid interfacial tension may be determined. We extract reasonable values for this quantity for four different liquid-solid combinations. Finally, we show that biaxial stresses in the capping elastic boundary can produce elliptical droplets with tuneable aspect ratio, and even droplets with nearly-square morphology using a suitable sample geometry.

\begin{figure}[]
     \includegraphics[width=\columnwidth]{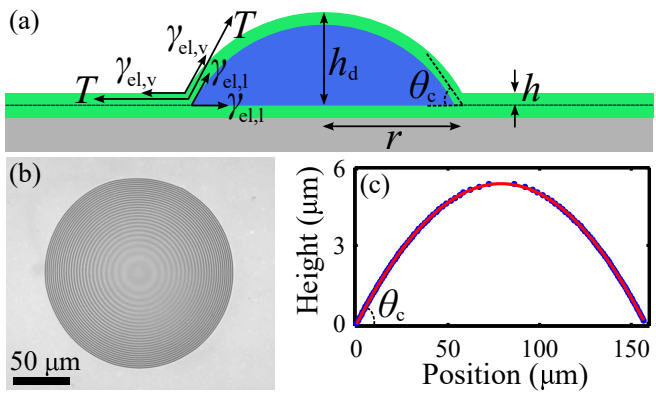}
\caption{(a) Schematic of a droplet capped by a taut elastic film and supported by a film of the same elastomer on a silicon substrate. The heuristic tension balance to determine the contact angle is shown. (b) Optical image under a red filter of the top view of a glycerol droplet capped by an Elastollan film with $\epsilon$ = 0.15 and $h$ = 570 nm. (c) Height profile of the capped droplet in (b) as a function of horizontal position. The solid curve is a circular cap fit from which we determine $\theta_\mathrm{c} = 7.8 \pm 0.2 ^\circ$. }
\label{fig1}
\end{figure}

Thin elastomeric films of Elastollan (elast) TPU 1185A (BASF)  and styrene-isoprene-styrene (SIS) triblock copolymer (14\% styrene content, Sigma-Aldrich) are prepared by spincoating out of cyclohexanone and toluene solutions respectively. Films are spun onto silicon wafers to create substrates and also onto freshly cleaved mica. The films are annealed at 150$^\circ$C (elast) or 100$^\circ$C (SIS) for 10 min to remove solvent and relax the polymer chains. The substrate films are of thickness 200 nm (elast) and 400 nm (SIS), as measured using ellipsometry (Accurion, EP3).  One edge of each substrate is brushed with an acetone-wetted cotton swab to remove the elastomer from this edge. Small droplets of glycerol (Caledon Laboratories Ltd.) or polyethylene glycol (PEG) with $M_n$ = 0.6 kg/mol (Sigma-Aldrich) are deposited onto substrates. These sessile droplets are capped with a thin film of the same elastomer as the substrate film (simplifying the boundary conditions) as follows: capping films are prepared on mica and transferred onto a home-built straining set-up following the protocol described in Ref.~\cite{Schulman2018b}. Using this apparatus, films are strained isotropically $\epsilon$, or biaxially with principal strains $\epsilon_\mathrm{low}$ and $\epsilon_\mathrm{high}$.  The substrate with droplets is then brought into contact with the strained capping film such that adhesive contact is formed between the two elastic films, thus completing the sample depicted in Fig.~\ref{fig1}(a). We study four different pairings of liquid and elastomer.  In the region where the substrate film was removed from the silicon using the cotton swab, we use ellipsometry to determine the thickness $h$ of the capping film. We employ films with $h$ between 150-1700 nm (Elastollan) and 550-3000 nm (SIS). 

After being capped, the droplets are circular with a contact radius $r$ when viewed from above. Only droplets which are not visibly pinned are measured. The droplets are in the range 30 $\mu$m $<r<$ 300 $\mu$m, which is large enough that evaporation can be ignored. Furthermore, these are much larger than the bulk elastocapillary length such that elastic substate films are not significantly deformed by capillarity and can be thought of as uncompliant substrates~\cite{Style2013a}. In addition, bending of the capping film is only relevant locally at the contact line and tension dominates the global picture~\cite{Andreotti2016b}. Due to the high tension in the capping film, the droplets are sufficiently flattened (to a height $h_\mathrm{d}$) to exhibit interference fringes when viewed under an optical microscope with a red ($\lambda$ = 632.8 nm) filter (Newport, 10LF10-633), as seen in Fig.~\ref{fig1}(b). From this interference pattern, the height profile of the droplet is determined. A height profile from a horizontal slice through the droplet in Fig.~\ref{fig1}(b) is shown by the data points in Fig.~\ref{fig1}(c). The profile is in excellent agreement with a circular cap fit, shown by the solid curve, from which the radius of curvature $R$ of the capped droplet is measured. If the droplets are large enough, it is also possible to directly image the droplet's profile by acquiring images from a side view, and fit this profile to a circle to extract $R$. We find good agreement between both techniques. The contact angle of the capped droplet $\theta_\mathrm{c}$ is evaluated using the geometric relation $\mathrm{sin}\theta_\mathrm{c} = r/R$. Several droplets are measured and the average $\theta_\mathrm{c}$ is determined. The contact angles are reproducible from one droplet to the next and independent of the size of the droplet over the range studied.

\begin{figure}[]
     \includegraphics[width=\columnwidth]{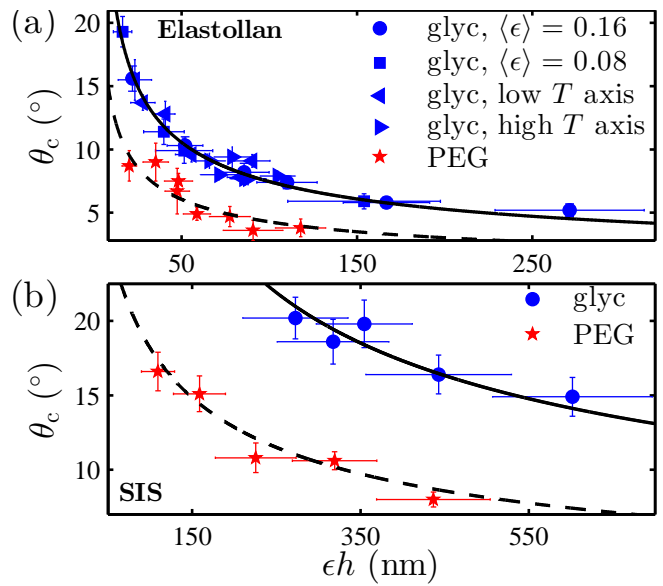}
\caption{Contact angle as a function of $\epsilon h$, which is proportional to tension, for two elastomers: (a) Elastollan and (b) SIS, and with two test liquids: glycerol (glyc) and PEG. The Elastollan/glycerol data includes sets where the strain is held constant, $\langle \epsilon \rangle = 0.16 \pm 0.03$ ($ \color{blue} \CIRCLE$) and  $\langle \epsilon \rangle = 0.08 \pm 0.02$ ($\color{blue} \blacksquare$), and only film thickness is changed. Plotted are also data attained from biaxial strain experiments along the low ($\color{blue} \LHD$) and high ($\color{blue} \RHD$) tension axes. Solid (glyc) and dashed (PEG) curves correspond to fits of Eq.~\ref{theta_c}, with $\gamma_\mathrm{el,l}/E$ as the only free parameter. Vertical error bars represent the standard deviation in contact angles measured for a sample, and horizontal error bars stem from uncertainties in thickness and strain. }
\label{fig2}
\end{figure}

Although elastomers are not Hookean over large strains,  we work exclusively with small strains ($\lessapprox$ 0.2), such that we can approximate a linear relationship between stress and strain~\cite{footnote:modulus}. As such, the mechanical tension in an isotropically strained film is $T = \frac{E\epsilon h}{1-\nu} \approx 2 E \epsilon h$, where $E$ is the Young's modulus of the elastomer and $\nu \approx 0.5$ is the Poisson ratio. Thus, the mechanical tension in a capping film can be varied by tuning film thickness or strain. In Fig.~\ref{fig2}(a), we plot the contact angle as a function of the product $\epsilon h$ for Elastollan with glycerol (glyc) droplets. Two data sets are shown wherein $h$ is varied while the isotropic strain is held constant: $\langle \epsilon \rangle = 0.16 \pm 0.03$ ($ \color{blue} \CIRCLE$) and  $\langle \epsilon \rangle = 0.08 \pm 0.02$ ($\color{blue} \blacksquare$). The data sets collapse on the same curve and show that contact angle decreases with $\epsilon h$. This observation is consistent with previous work which found dewetting rims of a liquid capped with an elastic film to be increasingly more flattened with higher $T$~\cite{Schulman2018b}. Our experiment is repeated with PEG droplets with strains in the range of 0.1-0.2 and variable film thicknesses, and the data is plotted in Fig.~\ref{fig2} ($ \color{red} \medblackstar$). Once again, the same trend is seen, but the two liquids, PEG and glycerol, do not collapse onto the same curve. In Fig.~\ref{fig2}(b), the two liquids are paired with SIS elastomer instead (again with strains in the range 0.1-0.2 and variable film thicknesses).

To understand these trends, we construct a balance of mechanical and interfacial tensions at the contact line akin to previous work~\cite{Nadermann2013,Hui2015,Schulman2016,Fortais2017,Schulman2017b}. We note that this balance is not truly a force balance, as the substrate exerts horizontal and vertical forces at the contact line as well. Rather, the balance represents changes in free energy for virtual motion of the contact line, in complete analogy with Young-Dupr\'{e}'s law. In the simplest model, we make the assumption that the tension in the capping film is not significantly altered by the deformation induced by the droplet (elastic membrane limit) and remains uniform throughout~\cite{landau1986}. This assumption is discussed later in more detail.  Although the balance is not truly a \emph{force} balance, we can make use of the analogy with Young-Dupr\'{e}'s law of partial wetting, and as such, the solution can be heuristically obtained by balancing the mechanical and interfacial tensions in the horizontal direction, as shown in Fig.~\ref{fig1}(a), where $\gamma$ represents interfacial tensions between elastomer ("el"), liquid ("l"), and vapour ("v"). Since the substrate is intentionally chosen to be the same material as the capping film, there is no interfacial tension between these. Motivated by previous work on these elastomers, we further assume that the interfacial tensions remain constant with strain (i.e. no Shuttleworth effect)~\cite{Schulman2018}. The horizontal  balance  gives $\mathrm{cos}\theta_\mathrm{c} = (T+\gamma_\mathrm{el,v}-\gamma_\mathrm{el,l})/(T+\gamma_\mathrm{el,v}+\gamma_\mathrm{el,l})$. Note that, just as for Young-Dupr\'{e}'s law, the same result is obtained by a free-energy minimization. Given measured modulus values $E_\mathrm{elast} = 13 \pm 2$ MPa and $E_\mathrm{SIS} = 1.1 \pm 0.2$~MPa~\cite{footnote:modulus}, we calculate that $300<T<7000$~mN/m for all our samples. Thus, we make the approximation that $T$ is much greater than any interfacial tension in the system, as interfacial tensions involving polymeric materials are typically $<50$~mN/m. Finally, we employ the small angle approximation (since all our contact angles are less than 20$^\circ$), and uncover a simple prediction for the contact angle:
\begin{equation}
\theta_\mathrm{c} = 2\sqrt{\frac{\gamma_\mathrm{el,l}}{T}} \approx \sqrt{2\frac{\gamma_\mathrm{el,l}}{  E \epsilon h}}.
\label{theta_c}
\end{equation}
Thus, $\theta_\mathrm{c}$ is fully determined by the mechanical tension and elastomer-liquid interfacial tension. To test the model, we fit Eq.~\ref{theta_c} to each data set in Fig.~\ref{fig2}, leaving $\gamma_\mathrm{el,l}/E$ as a free parameter. The fits are plotted in Fig.~\ref{fig2} as solid (glyc) and dashed (PEG) curves, and are in excellent agreement with the data. The fit values of $\gamma_\mathrm{el,l}/E$ are shown in Table~\ref{table1}. Using our measured values of $E$, we compute $\gamma_\mathrm{el,l}$. In addition, we have measured the Young's contact angle ($\theta_\mathrm{Y}$) of each solid/liquid combination tested, and the values are listed in Table~\ref{table1}. Using the Young-Dupr\'{e} equation, we can calculate the elastomer-vapour interfacial tension as $\gamma_\mathrm{el,v} = \gamma_\mathrm{l,v} \mathrm{cos} \theta_\mathrm{Y}+\gamma_\mathrm{el,l}$, where $\gamma_\mathrm{l,v}$ is found in the literature to be 63 mN/m~\cite{Lide2004} and 46 mN/m~\cite{Korosi1981} for glycerol and PEG. These interfacial energies (Table~\ref{table1}) are of typical magnitudes for interfacial tensions involving polymeric solids. In fact, $\gamma_\mathrm{el,v}$ for SIS is within error of that quoted in Ref.~\cite{Zuo2012}. Furthermore, the values of $\gamma_\mathrm{el,v}$ are determined twice (using the two liquids) for each solid, and are within error of each other, which further validates the simple model. 

\begin{table}[]
\centering
\caption{Interfacial tensions extracted from fitting to contact angle data.}
\label{table1}
\begin{ruledtabular}
\begin{tabular}{l |c c c c}
Solid/Liquid    & $\gamma_\mathrm{el,l}/E$  & $\gamma_\mathrm{el,l}$  & $\theta_\mathrm{Y}$   & $\gamma_\mathrm{el,v}$ \\
 & (nm) & (mN/m) & ($^\circ$) & (mN/m)   \\ [1ex]  \hline
Elast/glyc & $0.85 \pm 0.04$              & $11 \pm 2$                     & $67.8 \pm 0.8$           & $35 \pm 2$                     \\
Elast/PEG  & $0.28 \pm 0.08$              & $4 \pm 1$                      & $42.4 \pm 0.7$           & $38 \pm 1$                     \\
SIS/glyc        & $18 \pm 3$                   & $20 \pm 5$                     & $82.3 \pm 1.3$           & $29 \pm 5$                     \\
SIS/PEG         & $4.0 \pm 0.8$                & $5 \pm 1$                      & $54 \pm 1.2$             & $32 \pm 2$                    
\end{tabular}
\end{ruledtabular}
\end{table}

\begin{figure}[h]
     \includegraphics[width=\columnwidth]{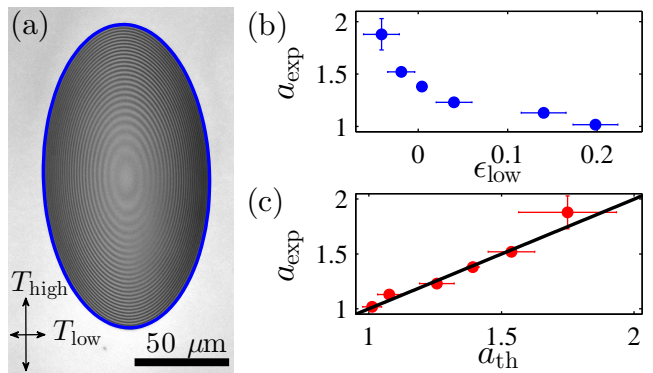}
\caption{(a) Optical image under red filter of an elliptical droplet aligned with the high tension direction in the capping film ($\epsilon_\mathrm{high}$ = 0.2, $\epsilon_\mathrm{low} $ = --0.04, $h$ = 586 nm). (b) Measured aspect ratio as a function of strain along the low tension axis. $\epsilon_\mathrm{high}$ is held constant at $\sim0.2$. (c) Measured aspect ratio plotted against the theoretically expected aspect ratio (Eq.~\ref{aspect_ratio}) given $\epsilon_\mathrm{high}$ and $\epsilon_\mathrm{low}$. The solid line represents the relationship $a_\mathrm{exp} = a_\mathrm{th}$. Horizontal error bars are due to uncertainties in the two strains. Vertical error bars are standard deviations in measured $a_\mathrm{exp}$. }
\label{fig3}
\end{figure}

To further characterize this system, we perform experiments with Elastollan and glycerol, wherein the droplets are capped by films with a biaxial tension with principal strains $\epsilon_\mathrm{high}$ and $\epsilon_\mathrm{low}$. In doing so, droplets assume an elongated shape (Fig.~\ref{fig3}(a)) which is well described by an ellipse whose major axis aligns with the high tension direction (plotted as a curve around the perimeter). To understand how the elliptical droplet's measured aspect ratio, $a_\mathrm{exp}$, varies with strain, we perform experiments where $\epsilon_\mathrm{high}\sim0.2$ and $-0.05<\epsilon_\mathrm{low}<0.2$. As seen in Fig.~\ref{fig3}(b), the aspect ratio decreases as $\epsilon_\mathrm{low}$ increases, and approaches unity in the limit of isotropic strain ($\epsilon_\mathrm{low} = \epsilon_\mathrm{high}$). 

We find that height profiles of the droplet along the high- and low-tension axes are well described by circular cap fits once again, and from these, we determine the contact angles along both axes ($\theta_\mathrm{c,high}$ and $\theta_\mathrm{c,low}$). Along each of the principal axes, the heuristic tension balance at the contact line, and the final prediction of Eq.~\ref{theta_c}, is identical to the isotropic case, but with $T$ replaced by $T_\mathrm{high}$ or $T_\mathrm{low}$ respectively. Applying Hooke's law, one can calculate the tension along the principal strain direction $i$ to be $T_i \approx \frac{1}{1-\nu^2}E(\epsilon_i+\nu\epsilon_j)h$ where $j$ is the orthogonal principal strain direction and for elastomers we can set $\nu \approx 0.5$. At this point, an effective strain can be defined for this direction $\epsilon_{\mathrm{eff},i} = \frac{1-\nu}{1-\nu^2}(\epsilon_i+\nu\epsilon_j)$, such that the isotropic expression is recovered for the biaxial case as well: $T_i = \frac{1}{1-\nu}E \epsilon_{\mathrm{eff},i}h \approx  2E\epsilon_{\mathrm{eff},i}h $. In this way, the elliptical droplet data for $\theta_\mathrm{c,high}$ ($\color{blue} \RHD$) and $\theta_\mathrm{c,low}$ ($\color{blue} \LHD$) is plotted against $\epsilon_{\mathrm{eff},i}h$ in Fig.~\ref{fig2}(a). In doing so, the biaxial strain data collapses onto the same curve as the isotropic strain data.

Since the droplet profiles along both principal directions are well fit to circular caps, the droplet height can be evaluated using the circular cap identity $h_{\mathrm{d},i} = r_i \mathrm{tan}(\theta_{\mathrm{c},i}/2) \approx r_i \theta_{\mathrm{c},i}/2$, where subscript $i$ once again denotes a principal direction and the small angle approximation was employed.  The droplet height must be the same for profiles taken along either principal direction ($h_\mathrm{d,high} = h_\mathrm{d,low}$). Therefore, the theoretical aspect ratio can be calculated as  $r_\mathrm{high}/r_\mathrm{low}$ to be
\begin{equation}
a_\mathrm{th} = \frac{\theta_\mathrm{c,low}}{\theta_\mathrm{c,high}} = \sqrt{\frac{T_\mathrm{high}}{T_\mathrm{low}}} = 
 \sqrt{\frac{\epsilon_\mathrm{high}+\nu\epsilon_\mathrm{low}}{\epsilon_\mathrm{low}+\nu\epsilon_\mathrm{high}}},
\label{aspect_ratio}
\end{equation}
where we have made use of Eq.~\ref{theta_c} and the Hookean relationships described above. For a
quantitative test of this result, we refer to Fig. \ref{fig3}(c), where all measurements of the aspect ratio $a_\mathrm{exp}$ are plotted against their predicted values $a_\mathrm{th}$, computed using Eq.~\ref{aspect_ratio} and the applied strains. Plotted in this way we find good agreement between theory and experiment (see solid line $a_\mathrm{exp} = a_\mathrm{th}$).

The theory outlined in this study relies on the assumption that the tension of the film is not significantly changed by the deformation induced by the droplet. To construct a comprehensive theoretical treatment to calculate the additional stresses created by this deformation, one may follow the methodology presented in articles by Davidovitch, Vella,  and co-workers, where the F\"{o}ppl-von K\'{a}rm\'{a}n equations are solved in the limit of vanishing bending contributions~\cite{Davidovitch2011,King2012,Schroll2013}. However, results we have presented suggest that our assumption is appropriate. Droplets under isotropic tension are well described as spherical caps, consistent with the notion that the tension in the deformed elastic film is isotropic and uniform (though not necessarily equal to the outside tension in the film) and is acting against a constant pressure within the capped liquid. The deformation of the capping film must introduce an additional strain, leading to an increased mechanical tension in this region. However, several pieces of empirical evidence suggest that this additional tension is negligible. The contact angles depend on the prepared tension, i.e. the product $\epsilon h$. If the tension was significantly modified during capping, one would expect the additional tension to depend individually on $h$ as well (through the stretching modulus $Eh$), and would disrupt the collapse of the data seen in Fig.~\ref{fig2}(a). Further evidence of the assumption's validity stems from the success and self-consistency of the model presented herein (success of predictions made by Eq.~\ref{theta_c} and Eq.~\ref{aspect_ratio}, sensible values of $\gamma_\mathrm{el,l}$ and $\gamma_\mathrm{el,v}$, and the collapse of biaxial and isotropic strain data in Fig.~\ref{fig2}(a)). However, we note that for our largest contact angles ($>15^\circ$), the assumption may begin to break down, and would certainly not be applicable for the significant deformations at large angles ($>30^\circ$).

\begin{figure}[h]
     \includegraphics[width=\columnwidth]{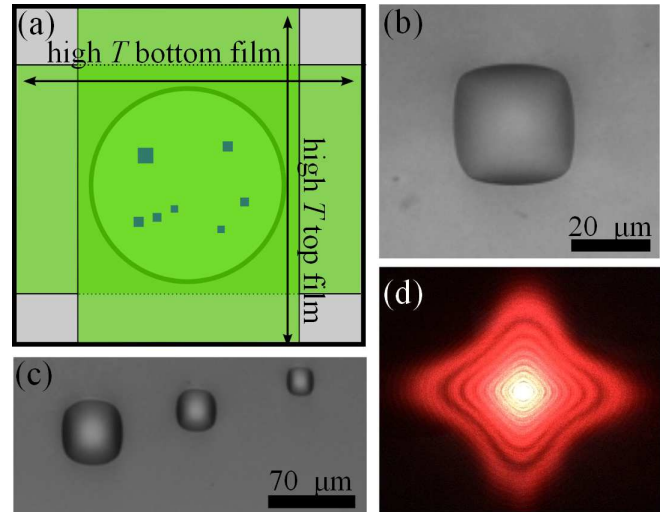}
\caption{(a) Top view schematic of the sample with droplets capped by biaxially stretched films on either side, all free-standing over the hole in the washer. (b)-(c) Optical images of droplets with square morphology. (d) Focal spot (with diffraction pattern) of a laser shone through a square droplet.}
\label{fig4}
\end{figure}

Motivated by previous work~\cite{Schulman2018b}, we perform an experiment with a unique combination of elastic boundary conditions, in the sample depicted in Fig.~\ref{fig4}(a). An Elastollan film with biaxial tension ($\epsilon_\mathrm{high} \sim 2$, $\epsilon_\mathrm{high} \sim -0.4$) is placed atop a washer with a circular hole. Glycerol droplets are deposited on the free-standing portion of this film. Finally, a second Elastollan film subjected to the same biaxial strains is placed on top, but with its high tension axis oriented perpendicular to the bottom film. Thus, the droplets are capped by an elastomer on either side, and the whole system is free-standing. As seen in Figs.~\ref{fig4}(b)-(c), droplets in such a sample assume a remarkable square morphology with the sides oriented along the principal strain directions. In this free-standing sample geometry (Fig.\ref{fig4}(a)), droplets act as lenses whose focal length and shape can be tuned. As a proof of principle, we shine a laser through a pinhole (150 $\mu$m in diameter) onto a square droplet. The resultant focal spot (with the diffraction pattern) is shown in Fig.~\ref{fig4}(d). As an elliptical droplet resembles a cylindrical lens producing a focal spot in the shape of a line, a square droplet creates a focal spot resembling a cross.

Here we have investigated the partial wetting of droplets capped by thin elastic films. When the tension in the elastic film is isotropic, droplets assume the shape of spherical caps which flatten with increasing tension. A balance of interfacial and mechanical tensions at the contact line made in analogy with Young-Dupr\'{e}'s law well describes the change in contact angle with tension with only one free parameter. From this free parameter, the elastomer-liquid interfacial tension -- a notoriously difficult quantity to measure -- can be determined. Finally, we show that elastic films as boundary conditions for partial wetting can produce droplets with novel morphologies. Droplets can be designed to have elliptical shapes with aspect ratios that depend on the strains in the elastic film, and it is even possible to generate droplets with square morphology using suitable choices of the elastic boundary conditions. Thus, elastic membranes can be used to create highly tunable liquid morphologies. 

The financial support by the Natural Science and Engineering Research Council of Canada the Joliot chair from ESPCI Paris is gratefully acknowledged. We thank A. Fortais for performing modulus measurements.


\begin{thebibliography}{34}%
\makeatletter
\providecommand \@ifxundefined [1]{%
 \@ifx{#1\undefined}
}%
\providecommand \@ifnum [1]{%
 \ifnum #1\expandafter \@firstoftwo
 \else \expandafter \@secondoftwo
 \fi
}%
\providecommand \@ifx [1]{%
 \ifx #1\expandafter \@firstoftwo
 \else \expandafter \@secondoftwo
 \fi
}%
\providecommand \natexlab [1]{#1}%
\providecommand \enquote  [1]{``#1''}%
\providecommand \bibnamefont  [1]{#1}%
\providecommand \bibfnamefont [1]{#1}%
\providecommand \citenamefont [1]{#1}%
\providecommand \href@noop [0]{\@secondoftwo}%
\providecommand \href [0]{\begingroup \@sanitize@url \@href}%
\providecommand \@href[1]{\@@startlink{#1}\@@href}%
\providecommand \@@href[1]{\endgroup#1\@@endlink}%
\providecommand \@sanitize@url [0]{\catcode `\\12\catcode `\$12\catcode
  `\&12\catcode `\#12\catcode `\^12\catcode `\_12\catcode `\%12\relax}%
\providecommand \@@startlink[1]{}%
\providecommand \@@endlink[0]{}%
\providecommand \url  [0]{\begingroup\@sanitize@url \@url }%
\providecommand \@url [1]{\endgroup\@href {#1}{\urlprefix }}%
\providecommand \urlprefix  [0]{URL }%
\providecommand \Eprint [0]{\href }%
\providecommand \doibase [0]{http://dx.doi.org/}%
\providecommand \selectlanguage [0]{\@gobble}%
\providecommand \bibinfo  [0]{\@secondoftwo}%
\providecommand \bibfield  [0]{\@secondoftwo}%
\providecommand \translation [1]{[#1]}%
\providecommand \BibitemOpen [0]{}%
\providecommand \bibitemStop [0]{}%
\providecommand \bibitemNoStop [0]{.\EOS\space}%
\providecommand \EOS [0]{\spacefactor3000\relax}%
\providecommand \BibitemShut  [1]{\csname bibitem#1\endcsname}%
\let\auto@bib@innerbib\@empty
\bibitem [{\citenamefont {de~Gennes}\ \emph {et~al.}(2008)\citenamefont
  {de~Gennes}, \citenamefont {Brochard-Wyart},\ and\ \citenamefont
  {Quere}}]{Gennes2008}%
  \BibitemOpen
  \bibfield  {author} {\bibinfo {author} {\bibfnamefont {P.}~\bibnamefont
  {de~Gennes}}, \bibinfo {author} {\bibfnamefont {F.}~\bibnamefont
  {Brochard-Wyart}}, \ and\ \bibinfo {author} {\bibfnamefont {D.}~\bibnamefont
  {Quere}},\ }\href {http://www.ulb.tu-darmstadt.de/tocs/114142769.pdf} {\emph
  {\bibinfo {title} {{Capillarity and Wetting Phenomena}}}}\ (\bibinfo
  {publisher} {Springer},\ \bibinfo {year} {2008})\BibitemShut {NoStop}%
\bibitem [{\citenamefont {Pericet-C\'{a}mara}\ \emph
  {et~al.}(2008)\citenamefont {Pericet-C\'{a}mara}, \citenamefont {Best},
  \citenamefont {Butt},\ and\ \citenamefont {Bonaccurso}}]{Best2008}%
  \BibitemOpen
  \bibfield  {author} {\bibinfo {author} {\bibfnamefont {R.}~\bibnamefont
  {Pericet-C\'{a}mara}}, \bibinfo {author} {\bibfnamefont {A.}~\bibnamefont
  {Best}}, \bibinfo {author} {\bibfnamefont {H.~J.}\ \bibnamefont {Butt}}, \
  and\ \bibinfo {author} {\bibfnamefont {E.}~\bibnamefont {Bonaccurso}},\
  }\href {\doibase 10.1021/la801862m} {\bibfield  {journal} {\bibinfo
  {journal} {Langmuir}\ }\textbf {\bibinfo {volume} {24}},\ \bibinfo {pages}
  {10565} (\bibinfo {year} {2008})}\BibitemShut {NoStop}%
\bibitem [{\citenamefont {Jerison}\ \emph {et~al.}(2011)\citenamefont
  {Jerison}, \citenamefont {Xu}, \citenamefont {Wilen},\ and\ \citenamefont
  {Dufresne}}]{Jerison2011}%
  \BibitemOpen
  \bibfield  {author} {\bibinfo {author} {\bibfnamefont {E.~R.}\ \bibnamefont
  {Jerison}}, \bibinfo {author} {\bibfnamefont {Y.}~\bibnamefont {Xu}},
  \bibinfo {author} {\bibfnamefont {L.~a.}\ \bibnamefont {Wilen}}, \ and\
  \bibinfo {author} {\bibfnamefont {E.~R.}\ \bibnamefont {Dufresne}},\ }\href
  {\doibase 10.1103/PhysRevLett.106.186103} {\bibfield  {journal} {\bibinfo
  {journal} {Phys. Rev. Lett.}\ }\textbf {\bibinfo {volume} {106}},\ \bibinfo
  {pages} {186103} (\bibinfo {year} {2011})}\BibitemShut {NoStop}%
\bibitem [{\citenamefont {Style}\ and\ \citenamefont
  {Dufresne}(2012)}]{Style2012}%
  \BibitemOpen
  \bibfield  {author} {\bibinfo {author} {\bibfnamefont {R.~W.}\ \bibnamefont
  {Style}}\ and\ \bibinfo {author} {\bibfnamefont {E.~R.}\ \bibnamefont
  {Dufresne}},\ }\href {\doibase 10.1039/c2sm25540e} {\bibfield  {journal}
  {\bibinfo  {journal} {Soft Matter}\ }\textbf {\bibinfo {volume} {8}},\
  \bibinfo {pages} {7177} (\bibinfo {year} {2012})}\BibitemShut {NoStop}%
\bibitem [{\citenamefont {Style}\ \emph
  {et~al.}(2013{\natexlab{a}})\citenamefont {Style}, \citenamefont
  {Boltyanskiy}, \citenamefont {Che}, \citenamefont {Wettlaufer}, \citenamefont
  {Wilen},\ and\ \citenamefont {Dufresne}}]{Style2013a}%
  \BibitemOpen
  \bibfield  {author} {\bibinfo {author} {\bibfnamefont {R.}~\bibnamefont
  {Style}}, \bibinfo {author} {\bibfnamefont {R.}~\bibnamefont {Boltyanskiy}},
  \bibinfo {author} {\bibfnamefont {Y.}~\bibnamefont {Che}}, \bibinfo {author}
  {\bibfnamefont {J.}~\bibnamefont {Wettlaufer}}, \bibinfo {author}
  {\bibfnamefont {L.~A.}\ \bibnamefont {Wilen}}, \ and\ \bibinfo {author}
  {\bibfnamefont {E.}~\bibnamefont {Dufresne}},\ }\href {\doibase
  10.1103/PhysRevLett.110.066103} {\bibfield  {journal} {\bibinfo  {journal}
  {Phys. Rev. Lett.}\ }\textbf {\bibinfo {volume} {110}},\ \bibinfo {pages}
  {066103} (\bibinfo {year} {2013}{\natexlab{a}})}\BibitemShut {NoStop}%
\bibitem [{\citenamefont {Marchand}\ \emph {et~al.}(2012)\citenamefont
  {Marchand}, \citenamefont {Das}, \citenamefont {Snoeijer},\ and\
  \citenamefont {Andreotti}}]{Marchand2012a}%
  \BibitemOpen
  \bibfield  {author} {\bibinfo {author} {\bibfnamefont {A.}~\bibnamefont
  {Marchand}}, \bibinfo {author} {\bibfnamefont {S.}~\bibnamefont {Das}},
  \bibinfo {author} {\bibfnamefont {J.~H.}\ \bibnamefont {Snoeijer}}, \ and\
  \bibinfo {author} {\bibfnamefont {B.}~\bibnamefont {Andreotti}},\ }\href
  {\doibase 10.1103/PhysRevLett.109.236101} {\bibfield  {journal} {\bibinfo
  {journal} {Phys. Rev. Lett.}\ }\textbf {\bibinfo {volume} {109}},\ \bibinfo
  {pages} {236101} (\bibinfo {year} {2012})}\BibitemShut {NoStop}%
\bibitem [{\citenamefont {Park}\ \emph {et~al.}(2014)\citenamefont {Park},
  \citenamefont {Weon}, \citenamefont {Lee}, \citenamefont {Lee}, \citenamefont
  {Kim},\ and\ \citenamefont {Je}}]{Park2014}%
  \BibitemOpen
  \bibfield  {author} {\bibinfo {author} {\bibfnamefont {S.~J.}\ \bibnamefont
  {Park}}, \bibinfo {author} {\bibfnamefont {B.~M.}\ \bibnamefont {Weon}},
  \bibinfo {author} {\bibfnamefont {J.~S.}\ \bibnamefont {Lee}}, \bibinfo
  {author} {\bibfnamefont {J.}~\bibnamefont {Lee}}, \bibinfo {author}
  {\bibfnamefont {J.}~\bibnamefont {Kim}}, \ and\ \bibinfo {author}
  {\bibfnamefont {J.~H.}\ \bibnamefont {Je}},\ }\href {\doibase
  10.1038/ncomms5369} {\bibfield  {journal} {\bibinfo  {journal} {Nat. Comm.}\
  }\textbf {\bibinfo {volume} {5}},\ \bibinfo {pages} {4369} (\bibinfo {year}
  {2014})}\BibitemShut {NoStop}%
\bibitem [{\citenamefont {Hui}\ and\ \citenamefont {Jagota}(2014)}]{Hui2014}%
  \BibitemOpen
  \bibfield  {author} {\bibinfo {author} {\bibfnamefont {C.-Y.}\ \bibnamefont
  {Hui}}\ and\ \bibinfo {author} {\bibfnamefont {A.}~\bibnamefont {Jagota}},\
  }\href@noop {} {\bibfield  {journal} {\bibinfo  {journal} {Pro. R. Soc. A}\
  }\textbf {\bibinfo {volume} {470}} (\bibinfo {year} {2014})}\BibitemShut
  {NoStop}%
\bibitem [{\citenamefont {Bostwick}\ \emph {et~al.}(2014)\citenamefont
  {Bostwick}, \citenamefont {Shearer},\ and\ \citenamefont
  {Daniels}}]{Bostwick2014}%
  \BibitemOpen
  \bibfield  {author} {\bibinfo {author} {\bibfnamefont {J.~B.}\ \bibnamefont
  {Bostwick}}, \bibinfo {author} {\bibfnamefont {M.}~\bibnamefont {Shearer}}, \
  and\ \bibinfo {author} {\bibfnamefont {K.~E.}\ \bibnamefont {Daniels}},\
  }\href@noop {} {\bibfield  {journal} {\bibinfo  {journal} {Soft Matter}\
  }\textbf {\bibinfo {volume} {10}},\ \bibinfo {pages} {7361} (\bibinfo {year}
  {2014})}\BibitemShut {NoStop}%
\bibitem [{\citenamefont {Shanahan}(1985)}]{Shanahan85}%
  \BibitemOpen
  \bibfield  {author} {\bibinfo {author} {\bibfnamefont {M.~E.}\ \bibnamefont
  {Shanahan}},\ }\href@noop {} {\bibfield  {journal} {\bibinfo  {journal} {The
  Journal of Adhesion}\ }\textbf {\bibinfo {volume} {18}},\ \bibinfo {pages}
  {247} (\bibinfo {year} {1985})}\BibitemShut {NoStop}%
\bibitem [{\citenamefont {Nadermann}\ \emph {et~al.}(2013)\citenamefont
  {Nadermann}, \citenamefont {Hui},\ and\ \citenamefont
  {Jagota}}]{Nadermann2013}%
  \BibitemOpen
  \bibfield  {author} {\bibinfo {author} {\bibfnamefont {N.}~\bibnamefont
  {Nadermann}}, \bibinfo {author} {\bibfnamefont {C.-Y.}\ \bibnamefont {Hui}},
  \ and\ \bibinfo {author} {\bibfnamefont {A.}~\bibnamefont {Jagota}},\ }\href
  {\doibase 10.1073/pnas.1304587110} {\bibfield  {journal} {\bibinfo  {journal}
  {Proc. Natl. Acad. Sci. U.S.A.}\ }\textbf {\bibinfo {volume} {110}},\
  \bibinfo {pages} {10541} (\bibinfo {year} {2013})}\BibitemShut {NoStop}%
\bibitem [{\citenamefont {Hui}\ \emph {et~al.}(2015)\citenamefont {Hui},
  \citenamefont {Jagota}, \citenamefont {Nadermann},\ and\ \citenamefont
  {Xu}}]{Hui2015}%
  \BibitemOpen
  \bibfield  {author} {\bibinfo {author} {\bibfnamefont {C.-Y.}\ \bibnamefont
  {Hui}}, \bibinfo {author} {\bibfnamefont {A.}~\bibnamefont {Jagota}},
  \bibinfo {author} {\bibfnamefont {N.}~\bibnamefont {Nadermann}}, \ and\
  \bibinfo {author} {\bibfnamefont {X.}~\bibnamefont {Xu}},\ }\href {\doibase
  10.1016/j.piutam.2014.12.013} {\bibfield  {journal} {\bibinfo  {journal}
  {Procedia IUTAM}\ }\textbf {\bibinfo {volume} {12}},\ \bibinfo {pages} {116}
  (\bibinfo {year} {2015})}\BibitemShut {NoStop}%
\bibitem [{\citenamefont {Hui}\ and\ \citenamefont {Jagota}(2015)}]{Hui2015b}%
  \BibitemOpen
  \bibfield  {author} {\bibinfo {author} {\bibfnamefont {C.-Y.}\ \bibnamefont
  {Hui}}\ and\ \bibinfo {author} {\bibfnamefont {A.}~\bibnamefont {Jagota}},\
  }\href@noop {} {\bibfield  {journal} {\bibinfo  {journal} {Soft Matter}\
  }\textbf {\bibinfo {volume} {11}},\ \bibinfo {pages} {8960} (\bibinfo {year}
  {2015})}\BibitemShut {NoStop}%
\bibitem [{\citenamefont {Schulman}\ and\ \citenamefont
  {Dalnoki-Veress}(2015)}]{Schulman2016}%
  \BibitemOpen
  \bibfield  {author} {\bibinfo {author} {\bibfnamefont {R.~D.}\ \bibnamefont
  {Schulman}}\ and\ \bibinfo {author} {\bibfnamefont {K.}~\bibnamefont
  {Dalnoki-Veress}},\ }\href {\doibase 10.1103/PhysRevLett.115.206101}
  {\bibfield  {journal} {\bibinfo  {journal} {Phys. Rev. Lett.}\ }\textbf
  {\bibinfo {volume} {115}},\ \bibinfo {pages} {206101} (\bibinfo {year}
  {2015})}\BibitemShut {NoStop}%
\bibitem [{\citenamefont {Liu}\ \emph {et~al.}(2016)\citenamefont {Liu},
  \citenamefont {Xu}, \citenamefont {Nadermann}, \citenamefont {He},
  \citenamefont {Jagota},\ and\ \citenamefont {Hui}}]{Liu2016}%
  \BibitemOpen
  \bibfield  {author} {\bibinfo {author} {\bibfnamefont {T.}~\bibnamefont
  {Liu}}, \bibinfo {author} {\bibfnamefont {X.}~\bibnamefont {Xu}}, \bibinfo
  {author} {\bibfnamefont {N.}~\bibnamefont {Nadermann}}, \bibinfo {author}
  {\bibfnamefont {Z.}~\bibnamefont {He}}, \bibinfo {author} {\bibfnamefont
  {A.}~\bibnamefont {Jagota}}, \ and\ \bibinfo {author} {\bibfnamefont {C.-Y.}\
  \bibnamefont {Hui}},\ }\href@noop {} {\bibfield  {journal} {\bibinfo
  {journal} {Langmuir}\ }\textbf {\bibinfo {volume} {33}},\ \bibinfo {pages}
  {75} (\bibinfo {year} {2016})}\BibitemShut {NoStop}%
\bibitem [{\citenamefont {Fortais}\ \emph {et~al.}(2017)\citenamefont
  {Fortais}, \citenamefont {Schulman},\ and\ \citenamefont
  {Dalnoki-Veress}}]{Fortais2017}%
  \BibitemOpen
  \bibfield  {author} {\bibinfo {author} {\bibfnamefont {A.}~\bibnamefont
  {Fortais}}, \bibinfo {author} {\bibfnamefont {R.~D.}\ \bibnamefont
  {Schulman}}, \ and\ \bibinfo {author} {\bibfnamefont {K.}~\bibnamefont
  {Dalnoki-Veress}},\ }\href@noop {} {\bibfield  {journal} {\bibinfo  {journal}
  {Eur. Phys. J. E}\ }\textbf {\bibinfo {volume} {40}},\ \bibinfo {pages} {69}
  (\bibinfo {year} {2017})}\BibitemShut {NoStop}%
\bibitem [{\citenamefont {Schulman}\ \emph {et~al.}(2017)\citenamefont
  {Schulman}, \citenamefont {Ledesma-Alonso}, \citenamefont {Salez},
  \citenamefont {Rapha{\"e}l},\ and\ \citenamefont
  {Dalnoki-Veress}}]{Schulman2017b}%
  \BibitemOpen
  \bibfield  {author} {\bibinfo {author} {\bibfnamefont {R.~D.}\ \bibnamefont
  {Schulman}}, \bibinfo {author} {\bibfnamefont {R.}~\bibnamefont
  {Ledesma-Alonso}}, \bibinfo {author} {\bibfnamefont {T.}~\bibnamefont
  {Salez}}, \bibinfo {author} {\bibfnamefont {E.}~\bibnamefont {Rapha{\"e}l}},
  \ and\ \bibinfo {author} {\bibfnamefont {K.}~\bibnamefont {Dalnoki-Veress}},\
  }\href@noop {} {\bibfield  {journal} {\bibinfo  {journal} {Phys. Rev. Lett.}\
  }\textbf {\bibinfo {volume} {118}},\ \bibinfo {pages} {198002} (\bibinfo
  {year} {2017})}\BibitemShut {NoStop}%
\bibitem [{\citenamefont {Style}\ \emph
  {et~al.}(2013{\natexlab{b}})\citenamefont {Style}, \citenamefont {Che},
  \citenamefont {Park}, \citenamefont {Weon}, \citenamefont {Je}, \citenamefont
  {Hyland}, \citenamefont {German}, \citenamefont {Power}, \citenamefont
  {Wilen}, \citenamefont {Wettlaufer},\ and\ \citenamefont
  {Dufresne}}]{Style2013}%
  \BibitemOpen
  \bibfield  {author} {\bibinfo {author} {\bibfnamefont {R.~W.}\ \bibnamefont
  {Style}}, \bibinfo {author} {\bibfnamefont {Y.}~\bibnamefont {Che}}, \bibinfo
  {author} {\bibfnamefont {S.~J.}\ \bibnamefont {Park}}, \bibinfo {author}
  {\bibfnamefont {B.~M.}\ \bibnamefont {Weon}}, \bibinfo {author}
  {\bibfnamefont {J.~H.}\ \bibnamefont {Je}}, \bibinfo {author} {\bibfnamefont
  {C.}~\bibnamefont {Hyland}}, \bibinfo {author} {\bibfnamefont {G.~K.}\
  \bibnamefont {German}}, \bibinfo {author} {\bibfnamefont {M.~P.}\
  \bibnamefont {Power}}, \bibinfo {author} {\bibfnamefont {L.~A.}\ \bibnamefont
  {Wilen}}, \bibinfo {author} {\bibfnamefont {J.~S.}\ \bibnamefont
  {Wettlaufer}}, \ and\ \bibinfo {author} {\bibfnamefont {E.~R.}\ \bibnamefont
  {Dufresne}},\ }\href@noop {} {\bibfield  {journal} {\bibinfo  {journal}
  {Proc. Natl. Acad. Sci. U.S.A.}\ }\textbf {\bibinfo {volume} {110}},\
  \bibinfo {pages} {12541} (\bibinfo {year} {2013}{\natexlab{b}})}\BibitemShut
  {NoStop}%
\bibitem [{\citenamefont {Alvarez}\ \emph {et~al.}(2018)\citenamefont
  {Alvarez}, \citenamefont {Bazilevs}, \citenamefont {Juanes},\ and\
  \citenamefont {Gomez}}]{Alvarez2018}%
  \BibitemOpen
  \bibfield  {author} {\bibinfo {author} {\bibfnamefont {J.~B.}\ \bibnamefont
  {Alvarez}}, \bibinfo {author} {\bibfnamefont {Y.}~\bibnamefont {Bazilevs}},
  \bibinfo {author} {\bibfnamefont {R.}~\bibnamefont {Juanes}}, \ and\ \bibinfo
  {author} {\bibfnamefont {H.}~\bibnamefont {Gomez}},\ }\href@noop {}
  {\bibfield  {journal} {\bibinfo  {journal} {Soft Matter}\ } (\bibinfo {year}
  {2018})}\BibitemShut {NoStop}%
\bibitem [{\citenamefont {Liu}\ \emph {et~al.}(2017)\citenamefont {Liu},
  \citenamefont {Nadermann}, \citenamefont {He}, \citenamefont {Strogatz},
  \citenamefont {Hui},\ and\ \citenamefont {Jagota}}]{Liu2017}%
  \BibitemOpen
  \bibfield  {author} {\bibinfo {author} {\bibfnamefont {T.}~\bibnamefont
  {Liu}}, \bibinfo {author} {\bibfnamefont {N.}~\bibnamefont {Nadermann}},
  \bibinfo {author} {\bibfnamefont {Z.}~\bibnamefont {He}}, \bibinfo {author}
  {\bibfnamefont {S.~H.}\ \bibnamefont {Strogatz}}, \bibinfo {author}
  {\bibfnamefont {C.-Y.}\ \bibnamefont {Hui}}, \ and\ \bibinfo {author}
  {\bibfnamefont {A.}~\bibnamefont {Jagota}},\ }\href@noop {} {\bibfield
  {journal} {\bibinfo  {journal} {Langmuir}\ }\textbf {\bibinfo {volume}
  {33}},\ \bibinfo {pages} {4942} (\bibinfo {year} {2017})}\BibitemShut
  {NoStop}%
\bibitem [{\citenamefont {Karpitschka}\ \emph {et~al.}(2016)\citenamefont
  {Karpitschka}, \citenamefont {Pandey}, \citenamefont {Lubbers}, \citenamefont
  {Weijs}, \citenamefont {Botto}, \citenamefont {Das}, \citenamefont
  {Andreotti},\ and\ \citenamefont {Snoeijer}}]{Karpitschka2016}%
  \BibitemOpen
  \bibfield  {author} {\bibinfo {author} {\bibfnamefont {S.}~\bibnamefont
  {Karpitschka}}, \bibinfo {author} {\bibfnamefont {A.}~\bibnamefont {Pandey}},
  \bibinfo {author} {\bibfnamefont {L.~A.}\ \bibnamefont {Lubbers}}, \bibinfo
  {author} {\bibfnamefont {J.~H.}\ \bibnamefont {Weijs}}, \bibinfo {author}
  {\bibfnamefont {L.}~\bibnamefont {Botto}}, \bibinfo {author} {\bibfnamefont
  {S.}~\bibnamefont {Das}}, \bibinfo {author} {\bibfnamefont {B.}~\bibnamefont
  {Andreotti}}, \ and\ \bibinfo {author} {\bibfnamefont {J.~H.}\ \bibnamefont
  {Snoeijer}},\ }\href@noop {} {\bibfield  {journal} {\bibinfo  {journal}
  {Proc. Natl. Acad. Sci. U.S.A.}\ }\textbf {\bibinfo {volume} {113}},\
  \bibinfo {pages} {7403} (\bibinfo {year} {2016})}\BibitemShut {NoStop}%
\bibitem [{\citenamefont {Martin}\ \emph {et~al.}(1997)\citenamefont {Martin},
  \citenamefont {Silberzan},\ and\ \citenamefont
  {Brochard-Wyart}}]{Martin1997}%
  \BibitemOpen
  \bibfield  {author} {\bibinfo {author} {\bibfnamefont {P.}~\bibnamefont
  {Martin}}, \bibinfo {author} {\bibfnamefont {P.}~\bibnamefont {Silberzan}}, \
  and\ \bibinfo {author} {\bibfnamefont {F.}~\bibnamefont {Brochard-Wyart}},\
  }\href@noop {} {\bibfield  {journal} {\bibinfo  {journal} {Langmuir}\
  }\textbf {\bibinfo {volume} {13}},\ \bibinfo {pages} {4910} (\bibinfo {year}
  {1997})}\BibitemShut {NoStop}%
\bibitem [{\citenamefont {Schulman}\ \emph
  {et~al.}(2018{\natexlab{a}})\citenamefont {Schulman}, \citenamefont {Niven},
  \citenamefont {Hack}, \citenamefont {DiMaria},\ and\ \citenamefont
  {Dalnoki-Veress}}]{Schulman2018b}%
  \BibitemOpen
  \bibfield  {author} {\bibinfo {author} {\bibfnamefont {R.~D.}\ \bibnamefont
  {Schulman}}, \bibinfo {author} {\bibfnamefont {J.~F.}\ \bibnamefont {Niven}},
  \bibinfo {author} {\bibfnamefont {M.~A.}\ \bibnamefont {Hack}}, \bibinfo
  {author} {\bibfnamefont {C.}~\bibnamefont {DiMaria}}, \ and\ \bibinfo
  {author} {\bibfnamefont {K.}~\bibnamefont {Dalnoki-Veress}},\ }\href@noop {}
  {\bibfield  {journal} {\bibinfo  {journal} {Soft matter}\ }\textbf {\bibinfo
  {volume} {14}},\ \bibinfo {pages} {3557} (\bibinfo {year}
  {2018}{\natexlab{a}})}\BibitemShut {NoStop}%
\bibitem [{\citenamefont {Farinha}\ \emph {et~al.}(2000)\citenamefont
  {Farinha}, \citenamefont {Winnik},\ and\ \citenamefont {Hahn}}]{Farinha2000}%
  \BibitemOpen
  \bibfield  {author} {\bibinfo {author} {\bibfnamefont {J.}~\bibnamefont
  {Farinha}}, \bibinfo {author} {\bibfnamefont {M.}~\bibnamefont {Winnik}}, \
  and\ \bibinfo {author} {\bibfnamefont {K.}~\bibnamefont {Hahn}},\ }\href@noop
  {} {\bibfield  {journal} {\bibinfo  {journal} {Langmuir}\ }\textbf {\bibinfo
  {volume} {16}},\ \bibinfo {pages} {3391} (\bibinfo {year}
  {2000})}\BibitemShut {NoStop}%
\bibitem [{\citenamefont {Andreotti}\ and\ \citenamefont
  {Snoeijer}(2016)}]{Andreotti2016b}%
  \BibitemOpen
  \bibfield  {author} {\bibinfo {author} {\bibfnamefont {B.}~\bibnamefont
  {Andreotti}}\ and\ \bibinfo {author} {\bibfnamefont {J.~H.}\ \bibnamefont
  {Snoeijer}},\ }\href@noop {} {\bibfield  {journal} {\bibinfo  {journal}
  {Europhys. Lett.}\ }\textbf {\bibinfo {volume} {113}},\ \bibinfo {pages}
  {66001} (\bibinfo {year} {2016})}\BibitemShut {NoStop}%
\bibitem [{foo()}]{footnote:modulus}%
  \BibitemOpen
  \href@noop {} {}\bibinfo {note} {We measure the modulus using tensile
  stress-strain measurements on fibers consisting of Elastollan and SIS with
  diameters between 5 and 40 microns. The stress-strain curves are linear
  within experimental error for strains up to 0.2.}\BibitemShut {Stop}%
\bibitem [{\citenamefont {Landau}\ and\ \citenamefont
  {Lifshitz}(1986)}]{landau1986}%
  \BibitemOpen
  \bibfield  {author} {\bibinfo {author} {\bibfnamefont {L.}~\bibnamefont
  {Landau}}\ and\ \bibinfo {author} {\bibfnamefont {E.}~\bibnamefont
  {Lifshitz}},\ }\href@noop {} {\emph {\bibinfo {title} {Theory of Elasticity,
  3rd}}}\ (\bibinfo  {publisher} {Butterworth-Heinemann, New York},\ \bibinfo
  {year} {1986})\BibitemShut {NoStop}%
\bibitem [{\citenamefont {Schulman}\ \emph
  {et~al.}(2018{\natexlab{b}})\citenamefont {Schulman}, \citenamefont {Trejo},
  \citenamefont {Salez}, \citenamefont {Rapha{\"e}l},\ and\ \citenamefont
  {Dalnoki-Veress}}]{Schulman2018}%
  \BibitemOpen
  \bibfield  {author} {\bibinfo {author} {\bibfnamefont {R.~D.}\ \bibnamefont
  {Schulman}}, \bibinfo {author} {\bibfnamefont {M.}~\bibnamefont {Trejo}},
  \bibinfo {author} {\bibfnamefont {T.}~\bibnamefont {Salez}}, \bibinfo
  {author} {\bibfnamefont {E.}~\bibnamefont {Rapha{\"e}l}}, \ and\ \bibinfo
  {author} {\bibfnamefont {K.}~\bibnamefont {Dalnoki-Veress}},\ }\href@noop {}
  {\bibfield  {journal} {\bibinfo  {journal} {Nat. Commun.}\ }\textbf {\bibinfo
  {volume} {9}},\ \bibinfo {pages} {982} (\bibinfo {year}
  {2018}{\natexlab{b}})}\BibitemShut {NoStop}%
\bibitem [{\citenamefont {Lide}(2004)}]{Lide2004}%
  \BibitemOpen
  \bibfield  {author} {\bibinfo {author} {\bibfnamefont {D.~R.}\ \bibnamefont
  {Lide}},\ }\href@noop {} {\emph {\bibinfo {title} {{CRC Handbook of Chemistry
  and Physics}}}}\ (\bibinfo  {publisher} {CRC Press},\ \bibinfo {year}
  {2004})\ pp.\ \bibinfo {pages} {6--154}\BibitemShut {NoStop}%
\bibitem [{\citenamefont {Korosi}\ and\ \citenamefont
  {Kovats}(1981)}]{Korosi1981}%
  \BibitemOpen
  \bibfield  {author} {\bibinfo {author} {\bibfnamefont {G.}~\bibnamefont
  {Korosi}}\ and\ \bibinfo {author} {\bibfnamefont {E.}~\bibnamefont
  {Kovats}},\ }\href {\doibase 10.1021/je00025a032} {\bibfield  {journal}
  {\bibinfo  {journal} {J. Chem. Eng. Data}\ }\textbf {\bibinfo {volume}
  {26}},\ \bibinfo {pages} {323} (\bibinfo {year} {1981})}\BibitemShut
  {NoStop}%
\bibitem [{\citenamefont {Zuo}\ \emph {et~al.}(2012)\citenamefont {Zuo},
  \citenamefont {Zheng}, \citenamefont {Zhao}, \citenamefont {Chen},
  \citenamefont {Yan}, \citenamefont {Ni},\ and\ \citenamefont
  {Wang}}]{Zuo2012}%
  \BibitemOpen
  \bibfield  {author} {\bibinfo {author} {\bibfnamefont {B.}~\bibnamefont
  {Zuo}}, \bibinfo {author} {\bibfnamefont {F.~F.}\ \bibnamefont {Zheng}},
  \bibinfo {author} {\bibfnamefont {Y.~R.}\ \bibnamefont {Zhao}}, \bibinfo
  {author} {\bibfnamefont {T.}~\bibnamefont {Chen}}, \bibinfo {author}
  {\bibfnamefont {Z.~H.}\ \bibnamefont {Yan}}, \bibinfo {author} {\bibfnamefont
  {H.}~\bibnamefont {Ni}}, \ and\ \bibinfo {author} {\bibfnamefont
  {X.}~\bibnamefont {Wang}},\ }\href@noop {} {\bibfield  {journal} {\bibinfo
  {journal} {Langmuir}\ }\textbf {\bibinfo {volume} {28}},\ \bibinfo {pages}
  {4283} (\bibinfo {year} {2012})}\BibitemShut {NoStop}%
\bibitem [{\citenamefont {Davidovitch}\ \emph {et~al.}(2011)\citenamefont
  {Davidovitch}, \citenamefont {Schroll}, \citenamefont {Vella}, \citenamefont
  {Adda-Bedia},\ and\ \citenamefont {Cerda}}]{Davidovitch2011}%
  \BibitemOpen
  \bibfield  {author} {\bibinfo {author} {\bibfnamefont {B.}~\bibnamefont
  {Davidovitch}}, \bibinfo {author} {\bibfnamefont {R.~D.}\ \bibnamefont
  {Schroll}}, \bibinfo {author} {\bibfnamefont {D.}~\bibnamefont {Vella}},
  \bibinfo {author} {\bibfnamefont {M.}~\bibnamefont {Adda-Bedia}}, \ and\
  \bibinfo {author} {\bibfnamefont {E.~A.}\ \bibnamefont {Cerda}},\ }\href
  {\doibase 10.1073/pnas.1108553108} {\bibfield  {journal} {\bibinfo  {journal}
  {Proc. Natl. Acad. Sci. U.S.A.}\ }\textbf {\bibinfo {volume} {108}},\
  \bibinfo {pages} {18227} (\bibinfo {year} {2011})}\BibitemShut {NoStop}%
\bibitem [{\citenamefont {King}\ \emph {et~al.}(2012)\citenamefont {King},
  \citenamefont {Schroll}, \citenamefont {Davidovitch},\ and\ \citenamefont
  {Menon}}]{King2012}%
  \BibitemOpen
  \bibfield  {author} {\bibinfo {author} {\bibfnamefont {H.}~\bibnamefont
  {King}}, \bibinfo {author} {\bibfnamefont {R.~D.}\ \bibnamefont {Schroll}},
  \bibinfo {author} {\bibfnamefont {B.}~\bibnamefont {Davidovitch}}, \ and\
  \bibinfo {author} {\bibfnamefont {N.}~\bibnamefont {Menon}},\ }\href
  {\doibase 10.1073/pnas.1201201109} {\bibfield  {journal} {\bibinfo  {journal}
  {Proc. Natl. Acad. Sci. U.S.A.}\ }\textbf {\bibinfo {volume} {109}},\
  \bibinfo {pages} {9716} (\bibinfo {year} {2012})}\BibitemShut {NoStop}%
\bibitem [{\citenamefont {Schroll}\ \emph {et~al.}(2013)\citenamefont
  {Schroll}, \citenamefont {Adda-Bedia}, \citenamefont {Cerda}, \citenamefont
  {Huang}, \citenamefont {Menon}, \citenamefont {Russell}, \citenamefont
  {Toga}, \citenamefont {Vella},\ and\ \citenamefont
  {Davidovitch}}]{Schroll2013}%
  \BibitemOpen
  \bibfield  {author} {\bibinfo {author} {\bibfnamefont {R.}~\bibnamefont
  {Schroll}}, \bibinfo {author} {\bibfnamefont {M.}~\bibnamefont {Adda-Bedia}},
  \bibinfo {author} {\bibfnamefont {E.}~\bibnamefont {Cerda}}, \bibinfo
  {author} {\bibfnamefont {J.}~\bibnamefont {Huang}}, \bibinfo {author}
  {\bibfnamefont {N.}~\bibnamefont {Menon}}, \bibinfo {author} {\bibfnamefont
  {T.}~\bibnamefont {Russell}}, \bibinfo {author} {\bibfnamefont
  {K.}~\bibnamefont {Toga}}, \bibinfo {author} {\bibfnamefont {D.}~\bibnamefont
  {Vella}}, \ and\ \bibinfo {author} {\bibfnamefont {B.}~\bibnamefont
  {Davidovitch}},\ }\href@noop {} {\bibfield  {journal} {\bibinfo  {journal}
  {Phys. Rev. Lett.}\ }\textbf {\bibinfo {volume} {111}},\ \bibinfo {pages}
  {014301} (\bibinfo {year} {2013})}\BibitemShut {NoStop}%
\end{thebibliography}

%

\end{document}